\documentclass[referee,traditabstract]{aa} 
\usepackage{graphicx}
\usepackage{natbib}
\begin{document}
\titlerunning{Generation of twist}
   \title{
   Generation of twist on magnetic flux tubes at the base of the solar
   convection zone}

   \author{H. Hotta
          \inst{1}
	  \and
	  T. Yokoyama\inst{1}
          }
	  
   	  \institute{
   	  Department of Earth and Planetary Science, University of
   	  Tokyo, 7-3-1 Hongo, Bunkyo-ku, Tokyo 113-0033, Japan}


 
  \abstract
  {
Using two-dimensional magnetohydrodynamics calculations, we investigate a
twist generation mechanism on a magnetic flux tube
at the base of the solar convection zone based on the idea of
Choudhuri, 2003, Sol. Phys., 215, 31 in which a toroidal magnetic field is
wrapped by a surrounding mean poloidal field. During generation of the twist,
the flux tube follows four phases. (1)
It quickly splits into two parts with vortex motions
rolling up the poloidal magnetic field. (2) Owing
to the
physical mechanism similar to that of the magneto-rotational instability, the
rolled-up poloidal field is bent and amplified. (3) The magnetic tension of
the disturbed poloidal magnetic field reduces the vorticity, and
the lifting force caused by vortical motion decreases.
(4) The flux tube gets twisted and begins to rise again without
 splitting.
Investigation of these processes is significant because it shows that
 a flux tube without any
 initial twist can rise to the surface in relatively weak poloidal fields.  
}
\maketitle

\section{Introduction}
Surface sunspots are the most prominent feature of solar
activity. It is thought that the magnetic flux is generated by the
dynamo action of the differential rotation at the base of the convection
zone
\citep{1955ApJ...122..293P,1995A&A...303L..29C,1999ApJ...518..508D,2010ApJ...709.1009H,2010ApJ...714L.308H}. 
The magnetic flux rises to the surface owing to the magnetic buoyancy and
generates sunspots on the surface.\par
In observational and theoretical assessments, this magnetic flux
is thought to be twisted.
On the surface, vector magnetic field observations have revealed that
the solar active regions have a statistically significant mean twist
\citep{1995ApJ...440L.109P,2001ApJ...549L.261P,2003ApJ...593.1217P}.
The quantity $\alpha\equiv\langle J_z/B_z\rangle$ is measured, where $J_z$, $B_z$,
and the parenthesis 
denote vertical electric current, vertical magnetic field, and spatial
average.
Although the measured $\alpha$ has large scatter, the quantity
has a latitudinal dependence of
$\alpha=-2.7\times10^{-10}\theta_\mathrm{deg}\ \mathrm{m^{-1}}$, where
$\theta_\mathrm{deg}$ is latitude in degree
\citep[see Fig. 2 of ][]{1998ApJ...507..417L}.
In addition, the theoretical (numerical) study of a rise in a flux
tube requires twist. \cite{1979A&A....71...79S} found that an untwisted
magnetic flux tube quickly splits into a pair of vortex tubes of opposite
circulation, which moves apart horizontally and ceases to rise. This
vortex is generated by the buoyancy gradient across the flux tube
cross-section. \cite{1996ApJ...472L..53M} reported that the sufficient
twist of a magnetic flux tube can suppress the splitting motion while
maintaining
coherency \cite[see also the review of][]{2009LRSP....6....4F}.\par
There are several possibilities for generating the twist. One
is that the dynamo process itself is responsible for it.
The second possibility is that the turbulent effect stemming from the
rotation on the
flux tube generates the twist, which is called the $\Sigma$-effect
\citep{1998ApJ...507..417L}. Another interesting explanation for the
origin of the twist is
given by \cite{2003SoPh..215...31C}. In his explanation, the rising
toroidal field is wrapped by a surrounding mean poloidal field and then
gets
twisted (see Fig. 4 of
his 2003 paper). \cite{2004ApJ...615L..57C} report that, based on this idea
and the Babcock-Leighton flux-transport dynamo model, the latitudinal
dependence of twist can be reproduced during most of the duration of the
solar
cycle. \cite{2006A&A...449..781C} modeled the evolution of the rising
magnetic flux tube with the accretion of the mean poloidal magnetic field by
solving the induction equation. They
report that without turbulent diffusivity the accreted poloidal flux
is confined in a sheath at the outer periphery of the rising tube. Using
plausible assumption and free parameters, they could obtain the value of
$\alpha$ that is comparable to the observational result.\par
In this paper, we investigate the generation mechanism of twist based on
the idea of \cite{2003SoPh..215...31C} in two-dimensional
magnetohydrodynamics (MHD) calculations using the parameters at the base of the
solar convection zone.
\section{Model}
We solve the compressive MHD equations in two-dimensional Cartesian
geometry ($x,z$), where $x$ and $z$ denote the horizontal and vertical
directions. The magnetic field has three components,
i.e., ${\bf B}=(B_x,B_y,B_z)$ even in two dimensions. Equations are
expressed as
\begin{eqnarray}
&& \frac{\partial \rho_1}{\partial
 t}=-\frac{1}{\xi^2}\nabla\cdot(\rho{\bf v}),\\
&& \frac{\partial }{\partial t}
  \left( \rho{\bf v}\right) = -\nabla\cdot
  \left[\rho{\bf vv}+\left(p_1+\frac{B^2}{8\pi}\right){\bf I}
  -\frac{\bf BB}{4\pi}\right]-\rho_1 g{\bf e_z},\\
&& \frac{\partial {\bf B}}{\partial t}=\nabla\times({\bf v}\times{\bf B}),\\
&&\frac{\partial s_1}{\partial t}=-{\bf v}\cdot\nabla s,\\
&& p_1 = p_0\left(\gamma\frac{\rho_1}{\rho_0}+\frac{s_1}{c_\mathrm{v}}\right)
\end{eqnarray}
where $\rho_1$, $p_1$, and $s_1$ denote the fluctuation of density,
pressure, and the specific entropy. ${\bf v}$, ${\bf B}$, $g$, and
$c_\mathrm{v}$ denote
fluid velocity, magnetic field, gravitational acceleration, and the heat
capacity at constant volume.
The divergence-free condition, i.e. $\nabla\cdot{\bf B}=0$, is
maintained numerically using the method introduced in
\cite{2002JCoPh.175..645D}. 
The medium is assumed to be an
inviscid perfect gas with a specific heat ratio $\gamma=5/3$.
The background values of density ($\rho_0$), pressure ($p_0$), and specific
entropy ($s_0$) are assumed to be in an adiabatic stratification with
constant gravitational acceleration as
\begin{eqnarray}
&&p_0 =
 p_\mathrm{b}\left[1-\nabla_\mathrm{ad}\frac{z}{H_\mathrm{b}}\right]^{1/\nabla_\mathrm{ad}},\\
&&\rho_0 =
 \rho_\mathrm{b}\left[1-\nabla_\mathrm{ad}\frac{z}{H_\mathrm{b}}\right]^{1/\nabla_\mathrm{ad}-1},\\
&&s_0 = c_\mathrm{v}\log\left(\frac{p_0}{\rho_0^\gamma}\right),\\
&&H_\mathrm{b} = \frac{p_\mathrm{b}}{\rho_\mathrm{b}g},
\end{eqnarray}
where $p_\mathrm{b}$, $\rho_\mathrm{b}$, and $H_\mathrm{b}$ are the
background pressure, density, and pressure scale height at the bottom
boundary. $\nabla_\mathrm{ad} = (\gamma-1)/\gamma$ denotes the adiabatic
temperature gradient.
 We use  the relation $\rho=\rho_0+\rho_1$, $p=p_0+p_1$, and
 $s=s_0+s_1$. The reduced speed of sound technique (hereafter
 RSST), which is introduced in \cite{2005ApJ...622.1320R} and
 \cite{2012A&A...539A..30H}, is used in this study, and $\xi$ denotes the
 ratio between the
 original and  reduced speed of sound. The investigation of the validity of
 RSST for the flux rising is given in Hotta et al. (in
 preparation). In the initial condition, the toroidal magnetic flux tube
 ($B_y$) is  set near the bottom boundary as
\begin{eqnarray}
&& B_y = \left\{ 
\begin{array}{ll}
B_\mathrm{t}\exp\left(-{\displaystyle \frac{r^2}{d^2}}\right) & \mathrm{where}\
 r<2d,\\
0 & \mathrm{where}\ r \geq 2d,
\end{array}
\right. \\
&& r^2 = (x-x_\mathrm{c})^2 + (z-z_\mathrm{c})^2,\\
&& B_\mathrm{t}=\sqrt{\frac{8\pi p_\mathrm{b}}{\beta_\mathrm{t} + 1}},
\end{eqnarray}
where $\beta_\mathrm{t}=1.6\times10^5$ is the plasma beta of the
toroidal field.
$(x_\mathrm{c},z_\mathrm{c})=(1.5H_\mathrm{b},0.15H_\mathrm{b})$, and
$d=0.03H_\mathrm{b}$ are the location and the radius of the flux tube.
Two cases with or without poloidal magnetic flux sheet are given in \S
\ref{result}.
A poloidal magnetic flux
 sheet ($B_x$) is set above the toroidal field as
\begin{eqnarray}\label{eq:poloidal}
&& B_x = B_\mathrm{p}\exp
\left[-
\left(\frac{z-z_\mathrm{p}}{d_\mathrm{p}}\right)^2\right],\\
&& z_\mathrm{p} = z_\mathrm{c}+2d+d_\mathrm{p},\\
&& B_\mathrm{p}=\sqrt{\frac{8\pi p_\mathrm{b}}{\beta_\mathrm{p} + 1}},
\end{eqnarray}
where $\beta_\mathrm{p}=10^{10}$ is the plasma beta for the
poloidal field, and $z_\mathrm{p}$ and $d_\mathrm{p}=0.08H_\mathrm{p}$ are
the location and thickness of the poloidal layer. The vertical magnetic
field is set as $B_z=0$. Figures \ref{magnetic}a and b show
pictures of the initial condition, where the fluctuation
of the pressure is set as $p_1 = -B^2/(8\pi)$, and $s_1=0$. 
The radius and
strength of the toroidal magnetic field is $d=1.8\times10^8\ \mathrm{cm}$ and
$B_\mathrm{}\sim 1\times10^5\ \mathrm{G}$
if we adopt the
parameters at the base of the solar convection zone;
i.e., $H_\mathrm{b}=6\times10^9\ \mathrm{cm}$,
$\rho_\mathrm{b}=0.2\ \mathrm{g\ cm^{-3}}$,
and the speed of sound at the
bottom $c_\mathrm{b}=2.2\times10^7\ \mathrm{cm\ s^{-1}}$. 
These values are plausible at the solar convection zone,
since the ``explosion'' process can amplify the magnetic field
up to $1\times10^5\ \mathrm{G}$ at the base \citep{2012ApJ...759L..24H},
and the magnetic field with this strength can reproduce the
observational result of the tilt angle of sunspot pairs at the surface,
i.e. Joy's law,
\citep{2011ApJ...741...11W}.
Owing to a lack of information on the strength and the distribution of
poloidal field from observation, 
we assumed
the thickness and the strength of the poloidal magnetic field layer as
$d=4.8\times10^{8}\ \mathrm{cm}$ and $B_\mathrm{p}\sim400\ \mathrm{G}$. \par
The simulation domain extends as
$(0,0)<(x/H_\mathrm{b},z/H_\mathrm{b})<(3,1.5)$.
The periodic boundary condition is used for all variables at
$x=0,\ 3H_\mathrm{b}$. At the top and bottom boundaries,
the stress-free and nonpenetrating boundary condition is adopted.
A symmetric boundary condition for the magnetic field and density, i.e. 
$\partial B_x/\partial z = \partial B_y/\partial z = 
\partial B_z/\partial z = 0$ and $\partial \rho_1/\partial z=0$, 
is used and entropy is fixed as $s_1=0$ at the top and bottom boundaries.
The adaptive mesh refinement (hereafter AMR) technique is also
 used, since in this study the magnetic flux is highly
 localized. A self-similar block-structured grid is adopted as the grid
 of the AMR hierarchy \citep{2007PASJ...59..905M}. In this study, the
 finest grid is equal to the grid using $1024\times512$ in uniform
 structure. The fourth-order,
 space-centered difference and fourth-order Runge-Kutta for the time
 integration are used \citep{2005A&A...429..335V}. The details of our AMR
 code will be given in our forthcoming paper. 
\section{Result and discussion}\label{result}
Figure \ref{normal} shows 
the result of the case without the mean poloidal field
($\beta_\mathrm{p}=\infty$)
 at $t=10^4H_\mathrm{b}/c_\mathrm{b}$.
 The magnetic flux tube quickly splits and ceases to rise. This
result is almost the same as previous studies
\citep{1979A&A....71...79S,1996ApJ...464..999L}.
Figure \ref{height} (red lines)
 shows the temporal evolution of the center of the splitted flux
 tube ($x_\mathrm{c},z_\mathrm{c}$) and the circulation $\Gamma$.
We define the center of a splitted flux tube ($x_\mathrm{c},z_\mathrm{c}$)
as the point where the magnetic
field has maximum value in the region $x>1.5H_\mathrm{b}$. The
circulation $\Gamma$  is defined as
\begin{eqnarray}
 \Gamma \equiv
  \int_{x_\mathrm{c}-d_\mathrm{c}}^{x_\mathrm{c}+d_\mathrm{c}}
  \int_{z_\mathrm{c}-d_\mathrm{c}}^{z_\mathrm{c}+d_\mathrm{c}}
  (\nabla\times{\bf v})_ydxdz,
\end{eqnarray}
where $d_\mathrm{c}=0.06H_\mathrm{b}$.
The lifting force caused by
circulation and horizontal motion ($F_\mathrm{lif}$)
 and the buoyancy ($F_\mathrm{buo}$) can be expressed as
\begin{eqnarray}
 &&F_\mathrm{lif} = -\rho V_x \Gamma, \\
 &&F_\mathrm{buo} = -g 
  \int_{x_\mathrm{c}-d_\mathrm{c}}^{x_\mathrm{c}+d_\mathrm{c}}
  \int_{z_\mathrm{c}-d_\mathrm{c}}^{z_\mathrm{c}+d_\mathrm{c}} \rho_1 dxdz,
\end{eqnarray}
where $V_x$ is the bulk horizontal velocity of the flux tube.
In the final stage,
the lifting force (directed downward) and the buoyancy (directed upward)
are almost balanced in the vertical direction. 
Since the flux tube rises
vertically in the first stage, it is accelerated horizontally by the horizontal
lifting force $\rho V_z \Gamma$,
where $V_z$ is the vertical bulk velocity of the
flux tube.
By this driven horizontal motion, a {\it downward} lifting force is
induced and must be balanced with the buoyancy force at a certain point.\par
Figures \ref{normal}c and d show the
distribution of angular velocity and angular momentum.
We define the center of the rotating motion at the point where vorticity
has
maximum value
(see the asterisk in Fig \ref{normal}b). The
result shows that the angular velocity increases radially, while the
angular momentum decreases.
We discuss the consequence of such distribution in \S 4.
\par
\begin{figure}[htbp]
 \centering
 \includegraphics[width=12cm]{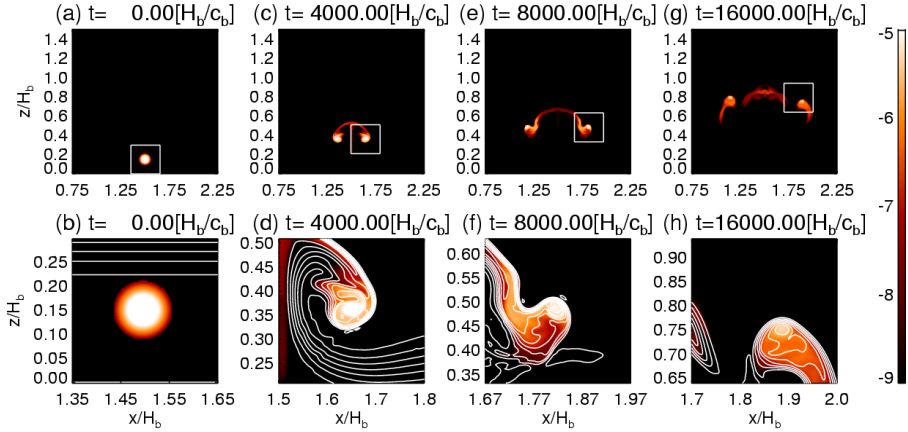}
 \caption{
 Temporal evolution of contours of 
toroidal magnetic field ($\log[(B_y/B_0)^2]$) in cases with a poloidal
 magnetic field,
where $B_0^2/(8\pi)=\rho_\mathrm{b}c_\mathrm {b}^2$. 
 The white rectangles in panels a, c, e, and g indicate the region of the
 panels b, d, f, and h, respectively.
 The white lines in panels b, d, f,
 and h show the poloidal magnetic field lines.
 \label{magnetic}}
\end{figure}
\begin{figure}[htbp]
 \centering
\includegraphics[width=12cm]{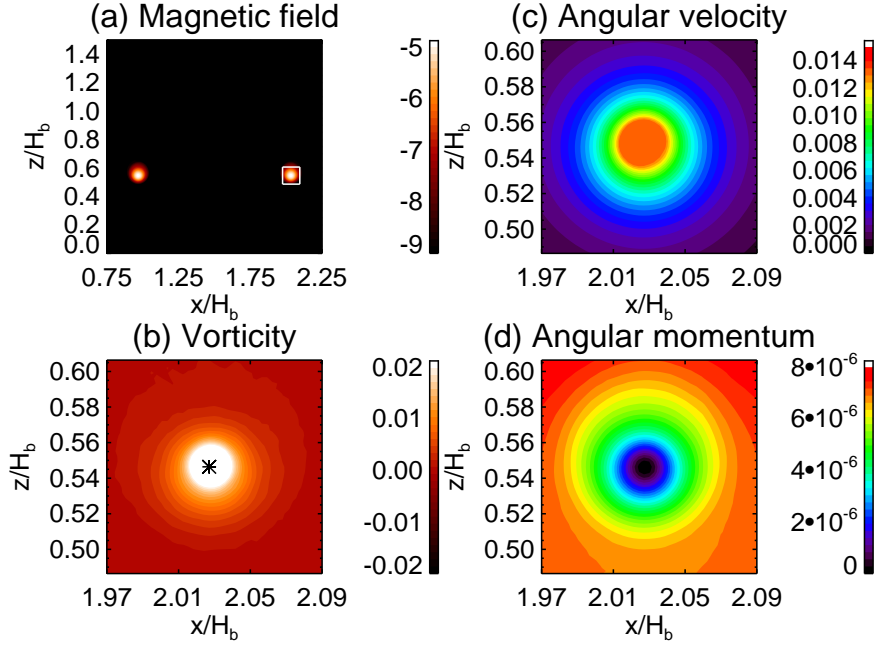}
 \caption{
 The results without poloidal field at
 $t=10^4H_\mathrm{b}/c_\mathrm{b}$. (a) The contour of toroidal magnetic
 field ($\log[(B_y/B_0)^2]$). 
 (b) The vorticity
 of $y$-direction
 in the white rectangle in panel a normalized by
 $c_\mathrm{b}/H_\mathrm{b}$. The point where vorticity has a maximum
 value is indicated with an asterisk. This point is defined as the origin
 of a circular geometry in order to calculate angular velocity and
 angular momentum. (c) The angular velocity normalized by
 $c_\mathrm{b}/H_\mathrm{b}$. (d) The angular momentum normalized by
 $\rho_\mathrm{b}H_\mathrm{b}c_\mathrm{b}$.
 \label{normal}}
\end{figure}
\begin{figure}[htbp]
 \centering
 \includegraphics[width=12cm]{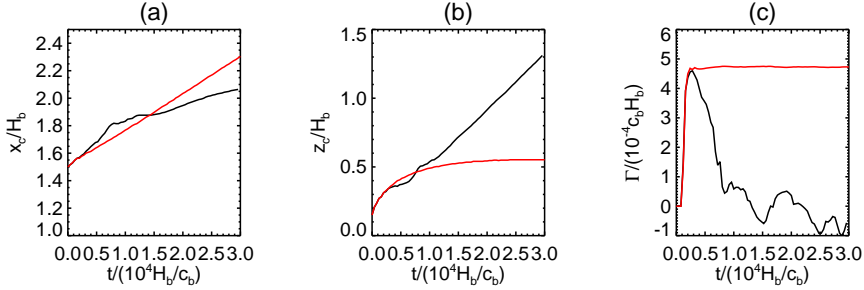}
 \caption{
Temporal evolution of the center of the flux tube, (a) $x_\mathrm{c}$,
 (b) $z_\mathrm{c}$, and (c) the circulation $\Gamma$.
 The black (red)
 line denotes the result with (without) the poloidal magnetic field.
 \label{height}}
\end{figure}
\begin{figure}[htbp]
 \centering
 \includegraphics[width=12cm]{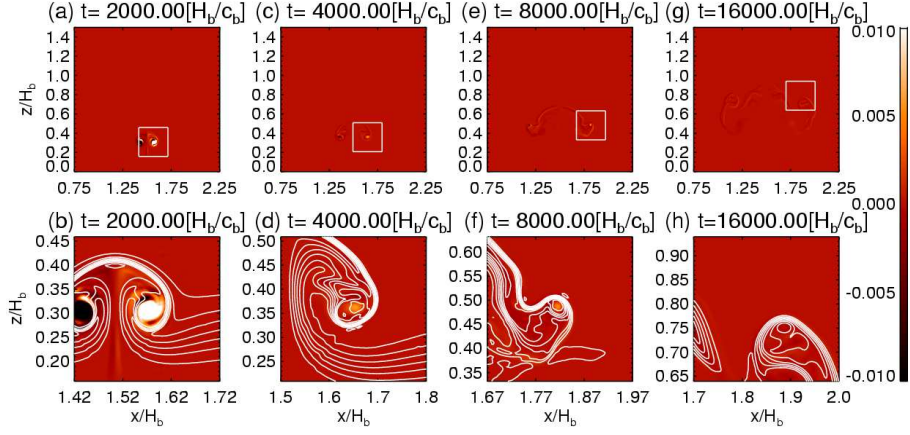}
 \caption{ Temporal evolution of contours of vorticity normalized by
 $c_\mathrm{b}/H_\mathrm{b}$ in the case with a poloidal field.
 The white rectangles in panels a, c, e, and g indicate the region of the
 panels b, d, f, and h, respectively.
 The white lines in panels b, d, f,
 and h show the poloidal magnetic field line.
 \label{vorticity}}
\end{figure}
\begin{figure}[htbp]
 \centering
 \includegraphics[width=12cm]{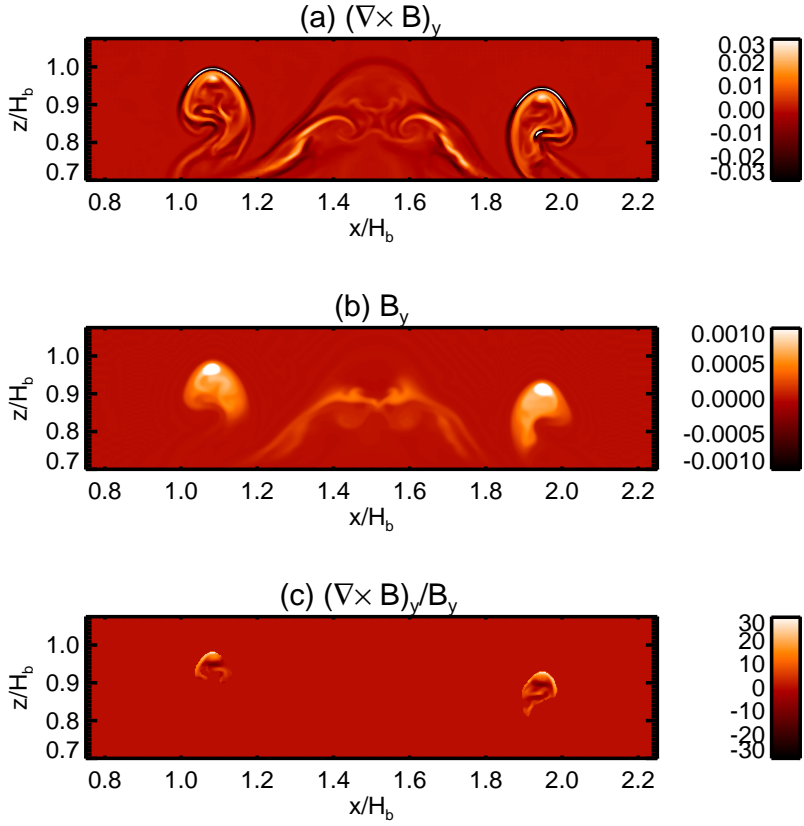}
 \caption{ Values in $t=20000H_\mathrm{b}/c_\mathrm{b}$. (a) The contour
 of $(\nabla\times{\bf B})_y$ normalized with
 $B_0/H_\mathrm{b}$. Positive value shows the clockwise poloidal
 fields.
 (b) The contour of Toroidal field $B_y$ normalized by $B_0$. (c) The
 contour of $(\nabla\times{\bf B})_y/B_y$ normalized by $H_\mathrm{b}^{-1}$.
 The value is only calculated in the place where the strength of
 toroidal field exceeds $6\times10^{-4}B_0$.
 \label{sign}}
\end{figure}

Figures \ref{magnetic} and \ref{vorticity} show the result of the case with
the mean poloidal field ($\beta_\mathrm{p}=10^{10}$)
\footnote{See also http://www-space.eps.s.u-tokyo.ac.jp/\%7ehotta/movie/twist.html}.
In the first stage, similar to the case without the poloidal field,
the flux tube splits
and moves horizontally with large vorticity (Figs. \ref{vorticity}a and
b). Then this vorticity gradually decreases
(Figs. \ref{vorticity}c, d, e, and f).
When the flux tube loses almost all circulation the magnetic
flux begins to rise again (Figs. \ref{magnetic}g, and h,
\ref{vorticity}g, and h). 
Figure \ref{height} (black line) also shows these processes, i.e.,
re-arisal occurs when circulation decreases around
$t=5000H_\mathrm{b}/c_\mathrm{b}$.
Figure \ref{magnetic}h shows that after this
process the magnetic flux gets twisted. Although it is not shown, these
twisted magnetic flux tubes rise to the upper boundary without
splitting in the final stage of this simulation.
We note that, although the flux tubes have vortices of different signs,
the signs of the generated twist on two flux tubes are the same
since the directions of the poloidal field and the toroidal field are
the same between two tubes.
 Figure \ref{sign}a, b, and c show the values of $(\nabla\times B)_y$, $B_y$,
 and $(\nabla\times B)_y/B_y$, respectively. The global structure of the
 poloidal field is shown
 clockwise, and the average of the values of the twist is positive near the
 core of the toroidal field in both flux tubes.
\par
\begin{figure}[htbp]
 \centering
 \includegraphics[width=12cm]{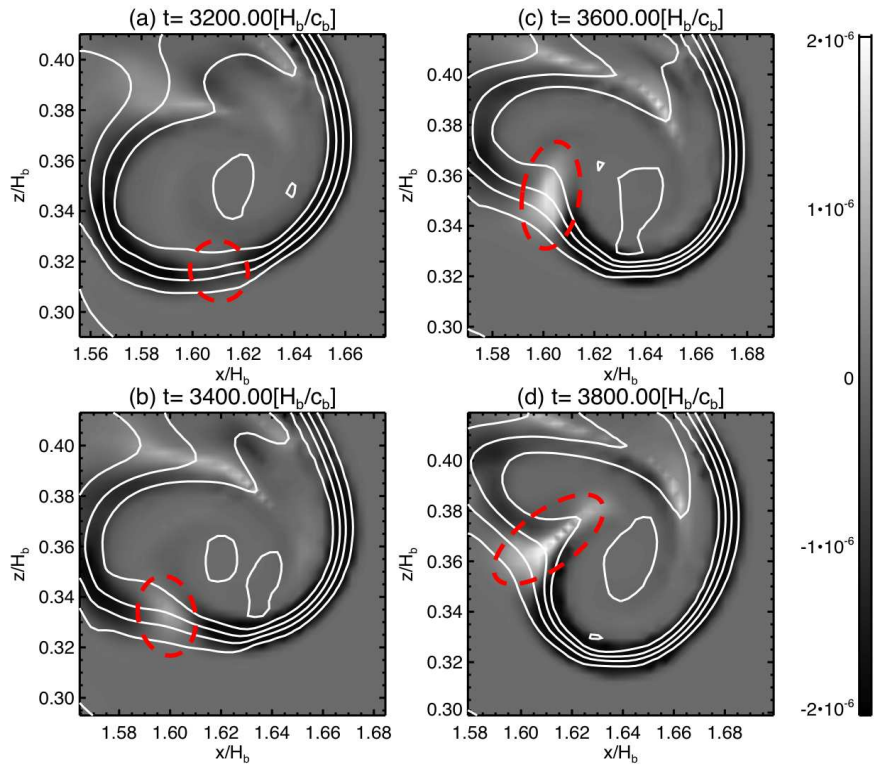}
 \caption{ Temporal evolution of magnetic tension
 ($T_\perp$: Eq. \ref{eq:tension}) normalized by $B_0^2/H_\mathrm{b}$
in the case with a poloidal field.
 Red circles indicate the location of the bending event.
 \label{tension}}
\end{figure}

We investigated the detailed process of the losing of circulation
in the the case with the initial poloidal
layer. Figure \ref{tension} shows the temporal
evolution of the magnetic tension perpendicular to the magnetic field line
($T_\perp$), i.e,
\begin{eqnarray}\label{eq:tension}
 T_\perp={\bf e}_\perp\cdot
\left[  \left({\bf B}\cdot\nabla\right) {\bf B} \right],
\end{eqnarray}
where ${\bf e}_\perp={\bf B}\times{\bf \omega}/(|{\bf B}||{\bf
\omega}|)$ is the unit vector
perpendicular to the magnetic field and the vorticity and ${\bf \omega}=\nabla\times{\bf v}$.
The poloidal field lines are
suddenly bent (see in Fig. \ref{tension}).
This type of sudden bending event on the magnetic field lines
occurs frequently at the point where the poloidal magnetic energy is
less than the energy of rotation.
Since the magnetic field is stretched and amplified during this
process,
the feedback by the magnetic tension efficiently suppresses circulating motion.
This can be seen in
Fig. \ref{tension}. The high positive value of the magnetic tension
indicates the feedback effect.
We recall that when the mean poloidal field does not exist, the angular
velocity increases and the angular momentum decreases radially
(Fig. \ref{normal}).
This configuration is similar to that of the magneto-rotational
instability
\citep{1960PNAS...46..253C}: the rolled-up poloidal magnetic
field can be bent
up to the interior of the flux tube. 
When the poloidal magnetic field has finite disturbance, the inner part
that is rotating faster loses the angular momentum due to the magnetic
tension, i.e., the interaction with the outer slow-rotating part. In this
case, the medium that loses angular momentum moves inward radially, and
the bending of the poloidal magnetic field increases.
Therefore the flux tube obtains the twist,
which distributes it over the tube's cross section without a
localization in the boundary layer.
We have determined that when the magnetic field is bent, the structure of circulation
is similar to that of the first case without the poloidal field;
i.e., the angular
velocity increases and the angular momentum decreases.\par
Our results seem to suggest that two flux tubes may emerge and form
active regions in different latitudes simultaneously. This probably does
not occur because of the modulation of the tube's emerging motion by the
thermal convection. \cite{2011ApJ...741...11W} have studied the rising
of the  magnetic flux
through the solar convection zone by using the thin-flux-tube
approximation with the effect of thermal convection and find that
the timing of emergence differs by more than one month between tubes from
different latitudes (see their Fig. 8) and that the emerging latitude is
also scattered typically over ten degrees (see their Fig. 11) with the
strength of a magnetic field of $4\times10^4$ G, which is the strength of our
splitted magnetic flux on their re-rising. Their results suggest that
our split pair tubes also emerge with these spatial and temporal
scattering offsets. Our splitted flux tubes separated about
$1H_\mathrm{p}$, which corresponds to five degrees in latitude and well below the
predicted scatter \citep[10 degrees,][]{2011ApJ...741...11W}
\section{Summary}
Using two-dimensional magnetohydrodynamics calculations, we investigated a
twist-generation mechanism on a magnetic flux tube
at the base of the solar convection zone. It is based on the idea of
\cite{2003SoPh..215...31C} in which a toroidal magnetic field is
wrapped by a surrounding mean poloidal field. During the generation of
the twist,
the flux tube follows four phases. (1)
The flux tube quickly splits into two parts with vortex motions
rolling up the poloidal magnetic field. (2) Because the
physical mechanism is similar to that of the magneto-rotational instability, the
rolled-up poloidal field is bent and amplified. (3) The magnetic tension of
the disturbed poloidal magnetic field reduces the vorticity, and
the lifting force caused by vortical motion decreases.
(4) The flux tube gets twisted and begins to rise again without splitting.
\par
With the use of the reduced RSST
\citep{2012A&A...539A..30H} and AMR, it became
possible to adopt
parameters that are acceptable at the base of the solar convection
zone, i.e., the radius of flux tube $d=1.8\times10^{8}\ \mathrm{cm}$ and
the strength of the toroidal field $B\sim1\times10^{5}\ \mathrm{G}$.
In this study, we use $400\ \mathrm{G}$ as the strength of the poloidal
field.
It is expected that a certain level of strength of such a poloidal field is
necessary for suppressing the circulation motion and the ensuing flux
tube's rising again.
The criterion for
rising again is an interesting issue.
The detailed parameter survey will be given in our forthcoming paper.

\begin{acknowledgements}
The authors wish to thank Tomoaki Matsumoto for helping us include AMR in our
 numerical code.
 This work was supported by Grant-in-Aid for JSPS Fellows.
We have greatly benefited from the proofreading/editing assistance
from the GCOE program.
\end{acknowledgements}

\end{document}